\documentclass[11p,reqno]{amsart}

\usepackage[T1]{fontenc}
\usepackage{graphicx}
\usepackage{amssymb,amsthm,amsmath,mathrsfs,bm,braket,marginnote}
\usepackage{enumerate}
\usepackage{appendix}
\usepackage[colorlinks=true, pdfstartview=FitV, linkcolor=blue, citecolor=blue, urlcolor=blue]{hyperref}
\usepackage{multirow}

\usepackage{pgf}
\usepackage{pgfplots}
\usepackage{tikz}
\usetikzlibrary{arrows,calc}
\usepackage{verbatim}
\usetikzlibrary{decorations.pathreplacing,decorations.pathmorphing}
\usepackage[numbers,sort&compress]{natbib}
\usepackage{dsfont}

\numberwithin{equation}{section}
\newtheorem{theorem}{Theorem}[section]

\theoremstyle{remark}
\newtheorem{remark}[theorem]{Remark}

 \reversemarginpar

\begin{document}

\title[Dyson's disordered linear chain]{Dyson's disordered linear chain from a random matrix theory viewpoint}

\author{Peter J. Forrester}
\address{School of Mathematical and Statistics, ARC Centre of Excellence for Mathematical and Statistical Frontiers, The University of Melbourne, Victoria 3010, Australia}
\email{pjforr@unimelb.edu.au}

\date{}

\dedicatory{}


\begin{abstract}
The first work of Dyson relating to random matrix theory, "The dynamics of a disordered linear chain'', is reviewed.
Contained in this work is an exact solution of a so-called Type I chain in the case of the disorder variables being 
given by a gamma distribution.  The exact solution exhibits a singularity in the density of states about
the origin, which has since been shown to be universal for one-dimensional tight binding models
with off diagonal disorder. We discuss this context and also point out some universal features of the weak disorder expansion
of the exact solution near the band edge. Further, a link between the exact solution,
and a tridiagonal formalism of anti-symmetric Gaussian
$\beta$-ensembles with $\beta$ proportional to $1/N$, is made.
\end{abstract}

\maketitle

\section{Introduction}
In the early 1960's Dyson, starting with the publication \cite{Dy62} and building on work of Wigner from the 1950's, developed
a theory of random matrices for applications to universal aspects of quantum spectra as determined by global symmetries.
For reference, we remark that these early works are conveniently reprinted and reviewed in a book edited by  Porter \cite{Po65}.
Whereas Wigner focussed on modelling the Hamiltonian using Hermitian random matrices, Dyson considered ensembles
of unitary matrices more fundamental due to there being a unique invariant measure; see Section I and the beginning of
Section II of \cite{Dy62}, and also the review \cite{DF17}. In addition to putting in place the mathematical framework, an
extensive theory was developed in relation to the statistical properties of the eigenvalues of the new ensembles.

It is no exaggeration to say these contributions of Dyson to random matrix theory and its applications are celebrated achievements.
Lesser known is the fact that these series of works were not the first time Dyson had use for random matrices, nor the first time
that he had the need to develop theory relating to random matrices in a pioneering fashion. The title to these claims goes instead to
Dyson's 1953 work ``The dynamics of a disordered linear chain'' \cite{Dy53}. 
From the viewpoint of foundational knowledge,  revisiting   \cite{Dy53} provides a valuable lesson in the methods and motivations
of random matrices.   And with Dyson's recent passing at age 96 on February 28th 2020, drawing attention
to \cite{Dy53} is also a contribution to paying tribute to his seminal contributions to the field generally.

There has been an earlier contextual review of  \cite{Dy53}, in the book of reprints with introductory text
on mathematical physics in one-dimension
written in the mid 1960's \cite[Ch.~2]{LM66} by Lieb and Mattis.
In addition to discussing follow up works from the original aims and objectives of \cite{Dy53}, this book also contains
reprints of the those papers. These follow ups are in relation to the properties of the disordered chain. A mathematical
follow up written in 1960, with the aim of putting all limiting procedures in \cite{Dy53} on a rigorous footing, can be found in the
work \cite{RB60} by a student of V.~Mar\v{c}enko, cited in the 1973 survey of Pastur \cite{Pa73} on the spectra of random
self adjoint operators. This latter reference also contains a discussion of \cite{Dy53}.

In Section \ref{S2}
an account is given of the salient content of \cite{Dy53}
from a random matrix theory viewpoint. Some subsequent refinements to
aspects of the working are covered in Section \ref{S3}, and also attention is drawn to universal features
of the exact solvable case found by Dyson. These are typically from the literature on localisation and the
one-dimensional Anderson model.
In Section \ref{S4}
a link between Dyson's solvable case,
and a tridiagonal formalism of anti-symmetric Gaussian
$\beta$-ensembles, with $\beta$ proportional to $1/N$ and for $N \to \infty$, is made.

\section{Overview of Dyson's paper}\label{S2}
\subsection{Coupled harmonic oscillators and tridiagonal matrices}
Dyson's Introduction in \cite{Dy53} makes it clear that his motivation was to present a mathematical model of a disordered system.
The particular choice made was a system of $N$ masses $\{ m_i \}_{i=1}^N$, confined to a line and each coupled to their nearest
neighbour by (fictitious) springs with corresponding springs constants $\{ K_i \}_{i=1}^N$, and obeying Hooke's law.
With free boundary conditions, the displacements from equilibrium of the positions $\{ u_i \}_{i=1}^N$ of each mass obey the coupled
set of Newton's equations
\begin{equation}\label{D1}
m_j {\ddot u}_j = K_j (u_{j+1} - u_j) + K_{j-1} (u_{j-1} - u_j).
\end{equation}
Here $K_0 = K_N = 0$ in keeping with free boundary conditions. The disorder is introduced by choosing the masses, or the spring constants,
or possibly a combination of both from a probability distribution function.

With $\mathbf a = [a_i]_{i=1}^N$, introduce the notation diag$\, \mathbf a$ for a matrix with entries given by $\mathbf a$ along the diagonal
and zero elsewhere. And with $\mathbf b = [b_i]_{i=1}^{N-1}$, introduce too the notation diag${}^+ \, \mathbf b$ ( diag${}^- \, \mathbf b$) for a matrix
with non-zero entries only on the first diagonal above (below) the main diagonal, with those entries given by $\mathbf b = [b_i]_{i=1}^{N-1}$.
To make use of this notation, set
\begin{align*}
&\mathbf u =  [u_i]_{i=1}^N, \qquad \boldsymbol{\alpha}_0 = [ - K_j/m_j - K_{j-1}/m_{j} ]_{j=1}^N \\
&   \boldsymbol{\alpha}_1 = [ K_j/ m_j ]_{j=1}^{N-1}, \qquad    \boldsymbol{\alpha}_{-1} = [ K_j/ m_{j+1} ]_{j=1}^{N-1}.
\end{align*}
We then have that the system (\ref{D1}) is equivalent to the second order matrix differential equation
\begin{equation}\label{D2}
 \ddot{\mathbf u} = \mathbf A \mathbf u, \qquad    \mathbf   A = {\rm diag} \, \boldsymbol{\alpha}_0 +   {\rm diag}^+ \,  \boldsymbol{\alpha}_1  +  {\rm diag}^- \,   \boldsymbol{\alpha}_{-1}.
 \end{equation}
 Separating variables by writing $\mathbf u = e^{i \omega t} \mathbf U$, where $\mathbf U$ is independent of $t$, shows the allowed values of
 $- \omega^2$ are given by the eigenvalues of the tridiagonal matrix $\mathbf A$.
 
 Instead of considering this eigenvalue problem, Dyson chose to first transform the second order system (\ref{D2}) into a first order system by
 changing variables $y_j = m_j^{1/2} u_j$ ($j=1,\dots,N$), then defining $\{ z_j \}_{j=1}^{N-1}$ by 
 \begin{equation}
 {\ddot z}_j = - \lambda_j^{1/2} y_j + \lambda_{j+1}^{1/2} y_{j+1},
  \end{equation}
  where
 \begin{equation}\label{D2a} 
\lambda_{2j-1} = K_j/ m_j, \qquad \lambda_{2j} = K_j/ m_{j+1}.
  \end{equation}
Introducing too
$$
\mathbf y = [ y_1 \: z_1 \: y_2 \: z_2 \cdots y_n]^T
$$
and with $\mathbf \Lambda$ the $(2N - 1) \times (2N - 1)$ anti-symmetric tridiagonal matrix
 \begin{equation}\label{2.5}
\mathbf \Lambda = {\rm diag}^+ \, [\lambda_j^{1/2} ]_{j=1}^{2N-1} -   {\rm diag}^- \, [\lambda_j^{1/2} ]_{j=1}^{2N-1},
  \end{equation}
  the second order matrix differential equation (\ref{D2}) is seen to be equivalent to the first order matrix
  differential equation
   \begin{equation}\label{D3}
   \dot{\mathbf y}  = \mathbf \Lambda  \, \mathbf y.
   \end{equation}
Separating variables by writing $\mathbf y = e^{i \omega t} \mathbf Y$, where $\mathbf Y$ is independent of $t$, shows the allowed values of
 $\omega$ are given by the $(N-1)$ positive eigenvalues of the matrix $i \mathbf A$, as well as the zero eigenvalue. The latter occurs due to the choice
 of free boundary conditions.
 
 Thus through either (\ref{D2}) or (\ref{D3}) Dyson was faced with the problem of quantifying the eigenvalue distribution for a (random) tridiagonal matrix.
 It was immediately realised that simplifying features could be expected in the limit $N \to \infty$. Thus define $M(\mu)$ as the proportion of frequencies $\{\omega_j\}$
 with $\omega_j^2 \le \mu$. Dyson hypothesised that for $N \to \infty$ \cite[Eq.~(10)]{Dy53}
  \begin{equation}\label{10}
  D(\mu) := {d M \over d \mu}
 \end{equation}
 is well defined, with $D(\mu)$ corresponding to the density of states  for the square eigenvalues.
 Under this assumption the function
  \begin{equation}\label{10s}   
  \Omega(x) := \lim_{N \to \infty} {1 \over (2 N - 1)} \sum_{j=1}^{N-1} \log (1 + x \omega_j^2),
  \end{equation}
referred to in \cite{Dy53} as the characteristic function of the chain,  is also well defined and can expressed in
terms of $D(\mu)$ according to  \cite[Eq.~(11)]{Dy53}
\begin{equation}\label{11}   
  \Omega(x)  = \int_0^\infty \log (1 + x \mu) D(\mu) \, d \mu.
  \end{equation}
  Moreover, it is noted that this can be inverted (differentiate and apply the Sokhotski--Plemelj --- also
  associated with Stieltjes--Perron --- formula) to deduce
 \cite[Eq.~(13)]{Dy53}    
\begin{equation}\label{13}    
D(1/x) = - x^2 \lim_{\epsilon \to 0^+} {1 \over \pi} {\rm Im} \, \Omega'(-x + i \epsilon).
 \end{equation}
 It is also noted that in the limit $x$ tends to $-z$ on the negative real axis from above, 
 ${\rm Im} \, \log (1 + x \mu) = 0$ ($i \pi$) for $z \mu < 1$ ($z \mu > 1)$ and thus
  \cite[Eq.~(12)]{Dy53}  
  \begin{equation}\label{12+}    
  {\rm Im} \, {1 \over \pi} \lim_{\epsilon \to 0^+} \Omega(-z + i \epsilon) =
  \int_{1/z}^\infty D(\mu) \, d \mu = 1 - M(1/z).
  \end{equation}
  Note the consistency between (\ref{12+}) and (\ref{13}).
 
 \subsection{A continued fraction formula for $\Omega (x)$}
 Dyson expands the logarithm in (\ref{10s}) to deduce
 \begin{equation}\label{16} 
 \sum_{j=1}^{N-1} \log (1 + x \omega_j^2) = \sum_{n=1}^\infty {(-1)^n \over n} x^n {\rm Tr} \, \mathbf  \Lambda^n.
  \end{equation}
  After some intricate combinatorial analysis of ${\rm Tr} \, \Lambda^n$, it is shown that for large $N$
  (\ref{16}) can be expressed in terms of the continued fraction  \cite[Eq.~(33)]{Dy53}   
  \begin{equation}\label{33}  
  \xi(a) = x \lambda_a/(1 + x \lambda_{a+1}/(1 + x \lambda_{a+2}/(1 + \cdots 
   \end{equation}
   Substituting in (\ref{10s}), this leads to the formula \cite[Eq.~(34)]{Dy53} 
   \begin{equation}\label{34}    
   \Omega(x) = \lim_{N \to \infty} {1 \over N} \sum_{a=1}^{2 N - 1} \log (1 + \xi(a)).
  \end{equation}  
  In \S \ref{S3.1} below, subsequent simplified derivations of (\ref{34}) will be given \cite{Be56, De56} which make use of
  algebraic rather than combinatorial properties of $\Lambda$.
  
  From the structure of (\ref{33}) and (\ref{34}), it is observed in \cite{Dy53} that the simplest type of disorder to impose is to
  choose $\{ \lambda_a \}$ from a common probability distribution. The coupled spring and masses system is then referred to 
  as a Type I disordered chain. With the probability density function (PDF) of the continued fraction (\ref{33}) denoted $F(\xi)$,
  (\ref{34}) then reads  \cite[Eq.~(46)]{Dy53} 
    \begin{equation}\label{46}  
    \Omega(x) = 2 \int_0^\infty F(\xi) \log (1 + \xi) \, d \xi.
  \end{equation}  
  
  An alternative type of disorder introduced in \cite{Dy53}, giving rise to what is termed a Type II disordered chain, is when each
  mass $m_j$ is an independent identically distributed random variable chosen with PDF $G(m)$, and with the spring constants
  all equal to the same value $K$. Then, from (\ref{D2a}) \cite[Eq.~(47)]{Dy53},
  $$
  \lambda_{2j-2} = \lambda_{2j-1} = K/ m_j    
 $$  
 so the random variables $\{ \lambda_j\}$ are constrained to be equal in pairs, while from (\ref{33}) $\xi(2j)$ and $\xi(2j-1)$ have
 different distributions. Defining \cite[Eq.~(48)]{Dy53}
 $$
 \eta_j = {1 \over \xi(2j)},
 $$
 and with $F(\eta)$ denote the corresponding PDF, manipulation of (\ref{34}) shows \cite[Eq.~(53)]{Dy53}
 \begin{equation}\label{53}   
    \Omega(x) =  \int_0^\infty d \eta \, F(\eta) \int_0^\infty d \tilde{m} \, G(\tilde{m}) \log \Big ( 1 + (1/ \eta) + x (K/\tilde{m}) \Big ).
  \end{equation} 
  
  As a distribution of masses of special interest, suppose      \cite[Eq.~(54)]{Dy53}
 \begin{equation}\label{54}    
 G(\tilde{m}) = p \delta (\tilde{m} - m) + (1 - p) \delta(\tilde{m} - M)
   \end{equation} 
   so that the chain consists of two masses $m, M$ with concentrations $p$ and $(1-p)$ respectively.
 Then (\ref{53}) reads    \cite[Eq.~(56)]{Dy53}
 \begin{equation}\label{56} 
 \Omega(x) = \int_0^\infty d \eta \, F(\eta) \bigg ( p \log \Big ( 1 + (1/\eta) + x (K/m) \Big ) 
 + (1 - p) \log \Big (  1 + (1/\eta) + x (K/M) \Big )     \bigg ).
 \end{equation}
 
 \subsection{A functional equation in the case of a Type I chain and an exact solution}
 The continued fraction (\ref{33}) obeys the functional equation   \cite[Eq.~(43)]{Dy53}
 \begin{equation}\label{43}
 \xi(a) = x \lambda_a/ (1 + \xi(a + 1)).
  \end{equation} 
  For a Type   I chain, the random variables $\lambda_a$ and $\xi(a+1)$ are uncorrelated and
  moreover $\xi(a)$ and $\xi(a+1)$ have the same distribution, leading to the equality in law
  between random variables $\xi$, and a combination of $\lambda$ and $\xi$,
  \begin{equation}\label{43a}
 \xi \overset{\rm d}{=}  x \lambda/ (1 + \xi).
  \end{equation}    
  Recalling now that the PDF for $\xi$ has been denoted $F(\xi)$ above (\ref{46}), and with the PDF
  for the distribution of $\lambda$ to be denoted $G(\lambda)$, we see that (\ref{43a}) implies
  \cite[equivalent to Eq.~(44)]{Dy53} 
  \begin{equation}\label{44}
  F(t) = \int_0^\infty d \lambda \, G(\lambda) \int_0^\infty d \xi \, F(\xi) \delta \Big ( t - {x \lambda \over 1 + \xi} \Big ).
  \end{equation}   
  
  With $\alpha, \kappa > 0$, suppose now     \cite[equivalent to eq.~(57)]{Dy53} 
 \begin{equation}\label{57}  
 G(\lambda) = {\kappa^\alpha \over \Gamma(\alpha)} \lambda^{\alpha - 1} e^{- \kappa \lambda}.
 \end{equation}
 Then, by taking the Mellin transform of both sides of (\ref{44}) it is straightforward to verify
 the fact that the solution of (\ref{44}) is  \cite[Eq.~(59)]{Dy53}
  \begin{equation}\label{59}  
 F(t) = F_\alpha(t) = {1 \over K_\alpha(x)} {t^{\alpha - 1} \over (1 + t)^\alpha} e^{- \kappa t/x},
  \end{equation} 
 where $ K_\alpha(x)$ is the normalisation given as an integral by
   \begin{equation}\label{59c}  
 K_\alpha(x) = \int_0^\infty {t^{\alpha - 1} \over (1 + t)^\alpha } e^{- \kappa t/x} \, dt.
  \end{equation}     
  Substituting in (\ref{46}) shows  \cite[Eq.~(60)]{Dy53}
     \begin{equation}\label{60}
     \Omega(x) = 2 {L_\alpha(x) \over  K_\alpha(x)},
 \end{equation}   
 where $ K_\alpha(x) $ is given by (\ref{59c}) and $L_\alpha(x)$ is given by
  \begin{equation}\label{61}   
L_\alpha(x) = \int_0^\infty {t^{\alpha - 1} \over (1 + t)^\alpha} \log (1 + t) e^{-\kappa t /x} \, dt.
 \end{equation}  
 
 \begin{remark}
 1.The use of the Mellin transform in relation (\ref{44}) is natural due to multiplicative nature of the noise;
 see \cite{Te20,CTT19} for recent developments. \\
 2. Denote by $\Gamma[\alpha,p]$ the gamma distribution with PDF proportional to $x^{\alpha - 1} e^{-px}$
 supported on $x > 0$. Denote by ${\rm K} [\alpha,\beta,p]$ the Kummer type II distribution with PDF proportional to
 $x^{\alpha - 1} e^{-p x}/ (1 + x)^{\alpha + \beta}$ supported on $x > 0$. A result attributed to Letac in
 an unpublished manuscript (see \cite[Remark 2.2]{HV16}) gives that with $X  \overset{\rm d}{=}  \Gamma[\alpha,p]$
 and  $Y  \overset{\rm d}{=}  {\rm K}[\alpha+\beta,-\beta,p]$,
\begin{equation}\label{L1}
{X \over 1 + Y}   \overset{\rm d}{=}     {\rm K}[\alpha,\beta,p].
\end{equation}  
With $\beta = 0$ this reduces to Dyson's result relating to (\ref{43a}).
\end{remark}

 In view of (\ref{13}), to compute the density of states, it is necessary to analytically continue both
 (\ref{59c})    and (\ref{61}) for negative $x$. This is done in  \cite[Appendix III]{Dy53}, and the result
 is substituted in the integrated form of (\ref{10})  \cite[equivalent to final equality in eq.~(12)]{Dy53} 
 \begin{equation}\label{12} 
 M(x) = 1 - \int_x^\infty D(\mu) \, d \mu.
  \end{equation}
  In the case that $\alpha = n \in \mathbb Z^+$, the explicit evaluation of (\ref{12}) was presented  \cite[Eq.~(63)]{Dy53}.
  Attention was drawn to the $x \to 0^+$ singularity \cite[consequence of eq.~(72)]{Dy53} 
 \begin{equation}\label{72}   
 M(x) \sim {c \over (\log x)^2}
 \end{equation}
 for some (explicit) $c$, now referred to as the Dyson spectral singularity.
 
 Attention was also drawn to the $\alpha \to \infty$ behaviour. In this limit, after setting $\kappa = \alpha$
 in (\ref{57}), the PDF for $\{\lambda_j\}$ has the asymptotic form  \cite[Eq.~(58)]{Dy53}
  \begin{equation}\label{58}  
  G(\lambda) \sim \Big ( {\alpha \over 2 \pi} \Big )^{1/2} e^{- \alpha (\lambda - 1)^2/2}
  \end{equation}
  of a Gaussian centred about $\lambda = 1$. To leading order each $\lambda_j$ is equal to $1$, and there is no
  disorder. It then follows from (\ref{43a}) that    \cite[Eq.~(37) with $\lambda = 1$]{Dy53}
\begin{equation}\label{37}   
\xi = {1 \over 2} \Big ( (1 + 4 x )^{1/2} - 1 \Big ),
  \end{equation}
  which substituted in (\ref{34}) gives  \cite[Eq.~(38)]{Dy53}
\begin{equation}\label{38}   
\Omega(x) = 2 \log \Big ( {1 \over 2} ( (1 + 4 x )^{1/2} + 1 ) \Big ).
\end{equation}
Substituting this in (\ref{13}) implies for the density of states   \cite[Eq.~(41) with $\lambda = 1$]{Dy53}
\begin{equation}\label{41}   
D(\mu) = \begin{cases} {1 \over \pi} {1 \over \sqrt{4  \mu - \mu^2}}, & \mu < 4  \\
0, & \mu > 4 \lambda.
\end{cases}
\end{equation}
A corollary of (\ref{41}), obtained by
substituting in (\ref{12}),  is that the integrated density of states for the chain with no disorder is \cite[Eq.~(74) with $\lambda = 1$]{Dy53}
\begin{equation}\label{74}   
M_\infty(x) = \begin{cases} {1 \over \pi} {\rm Arccos} (1 - x/2), & \mu < 4  \\
1, & \mu > 4 .
\end{cases}
\end{equation}
In relation to corrections to this behaviour due to disorder,
denote by $M_n(x)$ the integrated density of states in the case $\{\lambda_j\}$ are distributed with PDF
specialised to $\alpha = \kappa = n$. From his exact result, Dyson showed that for $n \to \infty$  \cite[Eq.~(75)]{Dy53}
\begin{equation}\label{75}   
M_n(x)  \sim 
 \begin{cases} {1 \over \pi}  {\rm Arcos} (1 - x/2) + {1 \over 2 \pi n} {1 \over (4/x - 1)^{1/2}}, & 0 < x < 4 \\
 1 - {\gamma \over \pi} \exp \Big ( - \gamma - 2n ( \sinh \gamma - \gamma)   \Big ), & x > 4 \\
 1 - \Big ( {1 \over \Gamma(1/3)} \Big )^2 \Big ( {12 \over n} \Big )^{1/3}, & x = 4,
 \end{cases}
 \end{equation}
 where $\gamma = {\rm Arcosh} \, ((x/2) - 1)$.
 
 \section{Some subsequent refinements}\label{S3}
 \subsection{Ratios of characteristic polynomials and Dyson's continued fraction}\label{S3.1}
 It was noted by Bellman \cite{Be56} and Dean \cite{De56} that Dyson's combinatorial derivation of (\ref{34})
 could be simplified by adopting an algebraic approach. For this purpose, in the notation of the paragraph including
 (\ref{D2}) introduce the general Hermitian tridiagonal matrix
 \begin{equation}\label{3.1} 
 \mathbf T_n = {\rm diag} \, [a_i]_{i=1}^n + {\rm diag}^+ \, [b_i]_{i=1}^{n-1} +  {\rm diag}^- \, [\bar{b}_i]_{i=1}^{n-1} .
  \end{equation}
  The corresponding (modified) characteristic polynomial is
 \begin{equation}\label{3.2}   
 P_n(y) = \det(  \mathbf  I_n - y  \mathbf  T_n) = \prod_{i=1}^n (1 - y \lambda_i^{(n)}),
 \end{equation}
 where $\{  \lambda_i^{(n)} \}$ are the eigenvalues of $\mathbf T_n$.
 By expanding $\det(  \mathbf  I_n - y  \mathbf T_n) $ along the final row, $\{ P_n(y) \}$ is seen to obey the three-term recurrence
 \begin{equation}\label{3.3}   
  P_n(y) =  (1 - y a_n) P_{n-1}(y) - y^2 | b_{n-1} |^2 P_{n-2}(y), \quad P_0(y) :=1.
 \end{equation}
 
 In terms of $r_n(y) := P_n(y)/ P_{n-1}(y)$, (\ref{3.3}) reads
  \begin{equation}\label{3.4}     
  r_n(y) = (1 - y a_n) - y^2 | b_{n-1} |^2 r_{n-1}(\lambda).
  \end{equation}
  With  $a_j = 0$ $(j=1,\dots, n)$ and upon the relabelling $b_{n-j} \mapsto b_j$, iteration of (\ref{3.4}) shows
   \begin{equation}\label{3.4a}     
 \lim_{n \to \infty}  r_{n+1-j}(y) =  1 - y^2 | b_j|^2/(1 - y^2 |b_{j+1}|^2)/(1 - y^2 |b_{j+2}|^2)/(1-\cdots.
   \end{equation}
   Furthermore, in terms of $\{ r_n(\lambda) \}$
 \begin{equation}\label{3.4b}   
   P_n(y) = \prod_{j=1}^n r_{n+1-j}(y).
 \end{equation} 
The significance of this setting is that in the case $n = 2N - 1$, with  diagonal entries given by $b_j = i \lambda_j$,
we have that $\mathbf T_n = \mathbf \Lambda$ and thus
 \begin{equation}\label{3.4c}  
P_n(y)  = \prod_{j=1}^{N-1} (1 - y^2 \omega_j^2 ).
 \end{equation} 
 Substituting (\ref{3.4c}) in (\ref{3.4b}) with  $-y^2$ replaced by $x$ , taking the logarithm and dividing by $(2N-1)$,
 then making use of (\ref{3.4a}) we see that (\ref{34}) is reclaimed.
 
 \subsection{Type II chain and the work of Schmidt}
 After separating variables as described below (\ref{D2}), the equations of motion (\ref{D1}) can be rearranged to
 read
  \begin{equation}\label{C1}
  U_{j+1} = \Big (1 + (K_{j-1}/K_j) - \omega^2 m_j/ K_j \Big ) U_j - (K_{j-1}/K_j) U_{j-1}.
   \end{equation} 
   To do this requires $K_j \ne 0$ for each $j=1,\dots,N$. This therefore excludes the free boundary conditions as used by Dyson,
   since then $K_N=0$ (recall the text below (\ref{D1})). A compatible  alternative  is to use fixed boundary
   conditions, specified by $U_0 = U_{N+1} = 0$. We remark that with (\ref{C1}) multiplied through by $K_j$ ($j=1,\dots,N$) free
   and fixed boundary conditions are indistinguishable.  Note too, following Schmidt \cite{Sc57}, that an equivalent way to write 
   (\ref{C1}) is as the $2 \times 2$ matrix recurrence
     \begin{equation}\label{C2}
     \begin{bmatrix} U_{j+1} \\ U_j \end{bmatrix} = \mathbf T_j
      \begin{bmatrix} U_{j} \\ U_{j-1}\end{bmatrix}, \qquad   \mathbf T_j = \begin{bmatrix} 1 + (K_{j-1}/K_j) - \omega^2 m_j/K_j & - K_{j-1}/K_j \\ 1 & 0 \end{bmatrix},
  \end{equation} 
  which implies the matrix product formula
  \begin{equation}\label{C2+}   
    \begin{bmatrix} U_{N+1} \\ U_N \end{bmatrix}    =  \mathbf T_N  \mathbf T_{N-1} \cdots \mathbf T_1    \begin{bmatrix} U_{1} \\ U_{0}\end{bmatrix}.
  \end{equation}    
   
  Iterating (\ref{C1}) starting with $U_0 = 0$ determines $\{ U_j \}_{j=1,2,\dots}$ up to an overall scalar factor $c$ say.
  We see  that $U_{j+1}$ is a polynomial of degree $j$ in $\omega^2$. Denoting the corresponding zeros by
   $\{ \mu_l^{(j)} \}_{l=1}^j$, this allows us to write
  \begin{equation}\label{C2a}  
 U_{j+1} =  c \prod_{l=1}^j (- m_l/K_l) ( \omega^2 - \mu_l^{(j)}).
 \end{equation} 
  To obtain a more explicit characterisation of   $\{ \mu_l^{(j)} \}_{l=1}^j$, in (\ref{C1}) change variables by
  writing
  \begin{equation}\label{C2b}  
  V_{j+1} = {1 \over \omega^{2j}} \prod_{l=1}^j \Big ( - {K_l \over m_l} \Big ) U_{j+1} , \qquad y = 1/\omega^2.
     \end{equation} 
 This gives
  \begin{equation}\label{C2c}  
  V_{j+1} =  1 - y \Big ( - {K_j \over m_j} -   {K_{j-1} \over m_j} \bigg ) V_j -
  y^2  \Big ( {K_{j-1}^2 \over m_{j-1} m_j} \Big ) V_{j-1}.
 \end{equation} 
 With $\mathbf A$ denoting the tridiagonal matrix specified in (\ref{D2}), and $\mathbf A_n$ denoting
 its top $n \times n$ submatrix, comparison with (\ref{3.3}) shows
   \begin{equation}\label{C2d}  
   V_{n+1} = \tilde{c} \det ( \mathbf I_n + y  \mathbf A_n),
   \end{equation}
 where $\tilde{c}$ is an arbitrary scalar. In particular, it follows that $\{ \mu_l^{(j)} \}_{l=1}^j$
 are equal to the nonzero eigenvalues of $-\mathbf  A_{j+1}$. Since $\mathbf  A_{N} = \mathbf A$
 it follows that for $U_{N+1} = 0$ as required by fixed boundary conditions, we must have
 that $-\omega^2$ in (\ref{C1}) corresponds to the eigenvalues of $\mathbf A$.
 This has been noted below
 (\ref{C2}) for the case of free boundary conditions.
 
The above theory implies that upon consideration of the ratios $\tilde{r}_n := U_n/ U_{n-1}$,
formulas equivalent to (\ref{3.4b}) and  (\ref{3.4c}) hold.  
Thus we have
 \begin{equation}\label{3.6}
 U_{N+1} =  c \prod_{l=1}^N (- m_l/K_l) ( \omega^2 - \omega_l^2) = \prod_{j=0}^N \tilde{r}_{N+1-j}.
 \end{equation}
However $\{\tilde{r}_n\}$ relate to the matrix $\mathbf A$, whereas 
$\{{r}_n\}$ relate to the anti-symmetric matrix $\mathbf \Lambda$ so they have
different distributions. In fact for a Type II chain,  characterised by all spring constants being equal, 
there are simplifications which result by considering $\{\tilde{r}_n\}$.

First, in the setting of a Type II chain,
it follows from (\ref{C1}) that
  \begin{equation}\label{C2e} 
 \tilde{r}_{j+1} = (2 - \omega^2 m_j / K) - 1/   \tilde{r}_{j}.
\end{equation} 
For $n$, large let $w(z) = w(z;\omega^2)$ denote the PDF for the distribution of $\tilde{r}_n$.
Proceeding as in the derivation of (\ref{44}), and specialising to the case of
a diatomic chain  as specified in (\ref{54}) for definiteness, we see that
$w(z)$ satisfies the functional equation \cite[equivalent to Eq.~(II,16)]{Sc57}
   \begin{equation}\label{C2g} 
   w(z) = p {1 \over z^2} w \Big (2 - {m \mu^2 \over K} - {1 \over z} \Big ) +
  (1 - p) {1 \over z^2} w \Big (2 - {M \mu^2 \over K} - {1 \over z} \Big ).
  \end{equation}
  
  Next, as a variant of Dyson's characteristic function (\ref{10s}) define
    \begin{align}\label{C2h} 
    \tilde{\Omega}(y^2)  & = \lim_{N \to \infty} {1 \over N} \sum_{j=1}^N \log ( \omega_j^2 - y^2 ) \nonumber \\
    & = \lim_{N \to \infty}    {1 \over N} \sum_{j=1}^N \log U_{N+1} \Big |_{\omega = y} +  \Big ( p \log m +
    (1 - p) \log M - \log K \Big ).
    \end{align}
  Comparison with (\ref{10s}) shows
  \begin{equation}\label{N1}
   \tilde{\Omega}(-1/z) = - \log z + \Omega(z).
     \end{equation}     
 In contrast to  (\ref{10s}), $ \tilde{\Omega}(y^2) $  is not real for positive real values of the argument $y^2$ ($x$ in  (\ref{10s})),
 since for $ \omega_j^2 < y^2 $
   \begin{equation}\label{C2k} 
 \log (  \omega_j^2 - y^2 ) = \log |  \omega_j^2 - y^2 | + i \pi.
  \end{equation}
 A useful consequence is that analogous to (\ref{12}), it follows
   \begin{equation}\label{C2p}  
   {\rm Im} \, {1 \over \pi}   \tilde{\Omega}(y^2)  =  M(y^2).
   \end{equation}

  From the second equality in (\ref{C2h}) and the definition of $w(z)$ it also follows that analogous to 
(\ref {46})
 \begin{equation}\label{C2m} 
    \tilde{\Omega}(y^2) = \int_{-\infty}^\infty  ( \log  z  )   w(z;y^2) \, dz.
  \end{equation}  
  Taking imaginary parts using (\ref{C2p}) then gives  \cite[Eq.~(II,26)]{Sc57}
  \begin{equation}\label{C2f} 
  M(y^2) = \int_{-\infty}^0 w(z;y^2) \, dz = {\rm Pr} \, \{ w(z;y^2) \le 0 \}.
 \end{equation} 
  We remark that due to this development of Schmidt to Dyson's pioneering work, (\ref{C2g}), along
  with (\ref{44}) in the case of the Type I chain, is nowadays typically referred to
  as an example of a Dyson--Schmidt equation for the stationary distribution of the corresponding
  stochastic sequences.   Also of note is that the final equality in (\ref{C2f}) is well suited to numerical
  approximation, whereas formalisms based on (\ref{12+}) require analytic continuation.
  
  \begin{remark}
  1. Consider the Type I disordered chain in Dyson's anti-symmetric tridiagonal matrix formulation. The characteristic
  polynomial $Q_n(y) = \det ( u \mathbf I_n - \mathbf \Lambda_n)$, where $\Lambda_n$ is the top $n \times n$ block
  of $\Lambda$ as specified by (\ref{2.5}) satisfies the recurrence
  $$
  Q_{n+1}(y) = y Q_n(y) - \lambda_n Q_{n-1}(y).
  $$
  Introducing the ratios $s_n = Q_n(y)/ Q_{n-1}(y)$, then writing $s_n = y (1 + \tilde{s}_n)$ we see
  that the random variable $\tilde{s}$ corresponding to the limiting distribution of $\tilde{s}_n$ must 
  satisfy the equality in law
   \begin{equation}\label{eL}
 \tilde{s}  \overset{\rm d}{=}  - { (\lambda / y) \over 1 + \tilde{s}}.
  \end{equation}    
 This is identical to (\ref{43a}) except that the positive parameter $x$ is now equal to the
 negative parameter $-1/y$. However, in relation to Dyson's exact solution in the case that
 the distribution of $\lambda$ is specified by (\ref{57}), having the parameter positive is an essential
 ingredient. In particular, no analogous exact solution is known in relation to (\ref{eL}). \\
 2.  Related to the above point is the simplification --- for example the existence of an exact
 solution --- which results by the consideration of $\Omega(z)$ (or  equivalently from (\ref{N1}) the
 consideration of $\tilde{\Omega}(-1/z)$)
 for $z$ positive and real, and then analytically continuing as required by (\ref{12+}) and (\ref{13}).
For related uses of this strategy, applied to multichannel models, see \cite{TBFM01,GT15}.
 \end{remark}
 
 Schmidt's reformulation of the second order difference system (\ref{C1}) in the matrix form (\ref{C2}) is
 significant as perhaps the first applied problem giving rise to a product of random matrices, as seen in
 (\ref{C2+}). In the case of the Type II chain, each matrix in (\ref{C2+}) is independent and identically
 distributed. Starting from the early 1960's, products of random matrices with
 independent and identically distributed elements attracted much attention in the mathematics
 literature, and many significant theoretical developments have followed \cite{FK60,Fu63,Os68,Ki73}.
 A key quantity in such studies as they relate to (\ref{C2+}) is
   \begin{equation}\label{Ly}
   \gamma := \lim_{N \to \infty} {1 \over N} \log \sqrt{ U_{N+1}^2 + U_N^2} = \lim_{N \to \infty} {1 \over N} \log |U_{N+1}|,
  \end{equation}  
  referred to as the Lyapunov exponent. In this setting it is usual to refer to the limiting PDF of
  $U_{N+1}/U_N$ as specifying an invariant measure.
  
  For Type II chains, it follows from the first equality in (\ref{3.6}) substituted in (\ref{Ly}) that
 \begin{equation}\label{Ly1} 
 \gamma = \int_0^\infty \log | \omega^2 - \mu| D(\mu) \, d\mu + {1 \over K} \langle \log |m| \rangle,
 \end{equation}  
 where $D(\mu)$ denotes the density of the squared singular eigenvalues; cf.~(\ref{11}). Nearly two
 decades after Dyson's work, it was understood by Thouless \cite{Th72} that through (\ref{Ly1}) there is a link between
 the density of states and the localisation length in a one-dimensional disordered system --- the latter
 being an interpretation of $1/\gamma$; see the review \cite{CTT13}. Due to this conceptual advance, 
 (\ref{Ly1}) is nowadays typically referred to as the Thouless formula, although some authors simultaneously cite
 both Dyson and Thouless; see e.g.~\cite{GID84}.
 
 \subsection{Type I chains near zero frequency}\label{S3.3}
 The natural discretisation of the one-dimensional Schr\"odinger equation
 $$
 \Big ( - {d^2 \over d x^2} + V(x) \Big ) \psi(x) = E \psi(x)
 $$
 on the integer lattice is
  \begin{equation}\label{1.1}
  - ( \psi_{n+1} + \psi_{n-1} - 2 \psi_n) + V_n \psi_n = E \psi_n.
 \end{equation} 
 This is to be compared with the difference equation (\ref{C1}) in the case of a type II chain
 \begin{equation}\label{1.2} 
 - ( U_{n+1} + U_{n-1} - 2 U_n) - {\omega^2 \over K} m_n U_n = 0.
 \end{equation} 
 While the discretisation of the Laplace operator $-d^2/d x^2$ is evident, there is no direct
 analogy between $\{E, V_n\}$ in (\ref{1.1}) and $\{\omega^2,m_n\}$ in (\ref{1.2}).
 
 On the other hand, consider Dyson's Type I chain in the anti-symmetric tridiagonal form
 (\ref{D3}). The corresponding equation for the eigenvalues and eigenvectors  can be written
 \begin{equation}\label{1.3} 
 i \lambda_{n-1}^{1/2} \phi_{n-1} - i \lambda_{n+1}^{1/2} \phi_{n+1} = \omega \phi_n,
  \end{equation}
  where $\{ \phi_n \}_{n=1}^{2 N - 1}$ are the components of the eigenvector.
  This is recognised as an example of the Schr\"odinger equation for the tight binding Hamiltonian
  (one-dimensional Anderson model) with random off diagonal elements and constant diagonal
  the latter being absorbed into the energy $E$ to give $\omega$ in (\ref{1.3}).
  
  A number of works have given consideration to the limiting $\omega \to 0^+$ form of the density of
  states as implied by (\ref{1.3}) for a general distribution of the non-negative random variable
  $\lambda_n$, assuming finite second moment; see for example \cite{TC76,ER78, Dh80,Bo89}. The
  conclusion of these works is that the singularity (\ref{72}) exhibited for the special distribution of
  $\lambda_n$ (\ref{57}) actually holds in the general case, and thus is a universal feature of both
  Dyson's Type I chain, and the one-dimensional Anderson model with off diagonal disorder.
  The same singularity is also seen in the density of states for the one-dimensional XY model with
  random coupling constants \cite{Sm70}, and the one-dimensional random
  mass Dirac Hamiltonian \cite{TH10}, both systems being related to the  Dyson's Type I chain.
  
  A closely related general behaviour can be seen from the solution of (\ref{1.3}) with $\omega = 0$
  \cite{TC76,  Zi82, CT98}.
  After iteration, and setting $\phi_1 = 1 $ for normalisation, we see that for $n$ even 
  $$
  \phi_{n+1} = \prod_{l=1}^{n/2} {\lambda_{2l-1} \over \lambda_{2l+1} }.
  $$
  According to the central limit theorem  $\log |  \phi_{n+1} |$ will, to leading order for
  large $n$, be proportional to $\sqrt{n}$, with the proportionality constant given in
  terms of the variance of $\lambda_n$. In contrast, typically for $\omega > 0$ the
  values of $\log |  \phi_{n+1} |$ obtained by iterating (\ref{1.3}) will be proportional to
  $n$. The modification of this conclusion in the case that the variance diverges has been
  the subject of the recent work \cite{KB20}; see too the earlier work
  \cite{BT08}.
  
  \subsection{Weak disorder limit}
  We know from (\ref{58}) that for $\alpha \to \infty$ the special PDF (\ref{57}) for the couplings
  $\{\lambda_n\}$ is to leading order a Gaussian centred at $\lambda = 1$ with variance $1/\alpha$.
  This circumstance, which perturbs about the chain with no disorder, is referred to as
  weak disorder. Systematic weak disorder expansion methods have been devised
  (see e.g.~\cite{CPV93} and references therein), typically specialised to the setting of the
  discrete Schr\"odinger equation (\ref{1.1}) and so not directly applicable to disordered
  chains. Nonetheless, comparison of the results which follow from Dyson's
  exactly solvable Type I chain  
   for large $\alpha$ 
 with results from the weak disorder expansion relating to
  (\ref{1.1}) (the pertinent ones are conveniently summarised in \cite[\S 2.2]{Lu19}),
  show a number of quantitative similarities.
  
  We consider first the Lyapunov exponent (\ref{Ly1}). A result of Thouless \cite{Th79}
  gives that in the weak disorder limit of  (\ref{1.1}), with the variance of $\{V_n\}$ equal to
  $1/\alpha$, the leading large $\alpha$ form for $|E| < 2$ is
   \begin{equation}\label{1.4} 
   \gamma \sim {1 \over 8 \alpha (1 - (E/2)^2)}.
  \end{equation}  
  For Type I chains, we see from the definition (\ref{Ly}) and (\ref{1.3}) that in terms of $D(\mu)$
  \begin{equation}\label{1.5} 
   \gamma = {1 \over 4} \int_0^\infty \log | \omega^2 - \mu| D(\mu) \, d \mu - {1 \over 2} \langle \log | \lambda| \rangle.
    \end{equation}   
 Making use of  (\ref{41}) it follows that with no disorder
   \begin{equation}\label{1.5-} 
   \gamma =  {1 \over 4 \pi} \int_0^4 {1 \over 4 \mu - \mu^2} \log | \omega^2 - \mu| \, d \mu = 0,
   \end{equation}
  where the final equality is valid for $\omega^2 < 4$; see e.g.~\cite[\S 1.4.2]{Fo10}.
  This fact could also be deduced directly from the definition (\ref{Ly}) since without disorder
  the components of the eigenvectors do not exponentially increase or decrease but rather
  oscillate, in keeping with the underlying coupled spring model having  all masses equal.
  
  Let $(\gamma)_1$ denote the term proportional to $1/\alpha$ in the large $\alpha$ expansion
  of $\gamma$, and similarly the meaning of $(\Omega(x))_1$. We see from (\ref{1.5}) and (\ref{11})
  that
  \begin{equation}\label{1.5+} 
   (\gamma)_1 =   {1 \over 4 \pi} \Big ( \lim_{\epsilon \to 0^+} {\rm Re} \, \Omega(-1/\omega^2 + i \epsilon) \Big )_1 - {1 \over 4}.
   \end{equation} 
   Making use of (\ref{60}) and workings in \cite[Appendix IV]{Dy53} gives, for $\alpha \in \mathbb Z^+$
   \begin{equation}\label{1.5a}  
    \lim_{\epsilon \to 0^+} {\rm Re} \, \Omega(-1/\omega^2 + i \epsilon) = 2 L_\alpha(-\omega^2)/ K_\alpha(-\omega^2),
    \end{equation} 
    where, with
   \begin{equation}\label{1.5b}     
   f(\xi,\omega) = \log \xi - \log (1 + \xi) + \xi \omega^2, \quad g_\alpha^{(0)}(\xi) = {1 \over \xi}, \quad
  g_\alpha^{(1)}(\xi) = {1 \over \xi} \log (1 + \xi),
   \end{equation} 
   we have \cite[Eqns.~(A.17) and (A.19)]{Dy53}
     \begin{align}
  L_\alpha(-\omega^2) & = \int_0^{-\infty}    
      g_\alpha^{(1)}(\xi ) e^{\alpha f(\xi,\omega)} \, d \xi \label{A.17} \\
 K_\alpha(-\omega^2) & = \int_0^{-\infty}    
      g_\alpha^{(0)}(\xi ) e^{\alpha f(\xi,\omega)} \, d \xi \label{A.19}      
   \end{align}
   In both (\ref{A.17}) and (\ref{A.19}) the contours of integration are to run along the
   upper half plane side of the negative real axis.
   
   Moreover, it is noted in \cite[Appendix IV]{Dy53} that for $0 < \omega^2 < 4$ there is a single
   saddle point in the upper half plane  \cite[Eq.~(A.21)]{Dy53}
     \begin{equation}\label{A.21} 
     \eta = {1 \over 2} \Big ( - 1 + i ( (4/\omega^2) - 1)^{1/2} \Big ).
  \end{equation}     
   It is noted too that by deforming the contours in (\ref{A.17}), (\ref{A.19}) to pass
   through this point at angle $3 \pi/ 4$, the large $\alpha$ asymptotic expansion follows by expanding
   the integrand about this point. Doing this, setting $f(\xi,\omega) = f(\xi)$ for notational convenience,
   and evaluating the corresponding Gaussian integrals gives
   \begin{align*}
     \lim_{\epsilon \to 0^+} {\rm Re} \, \Omega(-1/\omega^2 + i \epsilon) & =  - {1 \over 2} {1 \over \eta (1 + \eta)} {f'''(\eta) \over | f''(\eta) |^2}
     + {\rm Re} \, { 1 \over 2} {i \over (1 + \eta)^2 } {1 \over | f''(\eta) |} \\
     & = 1 + {1 \over 2 ((4/\omega)^2 - 1)}.
     \end{align*}
     Now substituting in (\ref{1.5+}) gives
    \begin{equation}\label{1.5c}   
    (\gamma)_1 = {1 \over 8 ( (4/\omega)^2 - 1)}.
    \end{equation}
    Comparing with (\ref{1.4}) shows the functional forms agree for $E = \omega \to 2^-$.     
    
    For the discrete Schr\"odinger equation (\ref{1.1}) it is known \cite{FL60,Ha65,DG84,IRT98,CLTT13} that
    in the limit $E \to 2$ the leading weak disorder expansion of the Lyapunov exponent obeys the scaling law
    in terms of  Airy functions
    \begin{equation}\label{1.5d}      
   \gamma \sim   \Big ( {1 \over 2 \alpha} \Big )^{1/3} F \Big ((2 \alpha)^{2/3} ( |E| - 2) \Big ),
    \end{equation}
    where
   \begin{equation}\label{1.5e}   
   F(x) = {{\rm Ai}(x)    {\rm Ai}'(x)  + {\rm Bi}(x)    {\rm Bi}'(x) \over ({\rm Ai}(x) )^2 + ({\rm Bi}(x) )^2} =  {\rm Re} \, e^{-2 \pi i /3} { {\rm Ai}' ( e^{- 2 \pi i /3} x) \over  {\rm Ai} ( e^{- 2 \pi i /3} x) }.
   \end{equation} 
   Precisely this scaling function was obtain by Smith \cite[Eq.~(4.14)]{Sm70} in the case of Dyson's exactly solvable Type I chain with $\alpha \to \infty$.
   It was found by extending the asymptotic analysis of Dyson to uniformly account for  there being two coalescing saddle
   points as $\omega \to 2$.
   
   \begin{remark}
   It is noted in \cite[\S 4.1]{CLTT13} that
   \begin{equation}\label{1.5f}  
   {\rm Im} \, e^{-2 \pi i /3} { {\rm Ai}' ( e^{- 2 \pi i /3} x) \over  {\rm Ai} ( e^{- 2 \pi i /3} x) } = {1 \over \pi ( ({\rm Ai}\,(x))^2 +  ({\rm Bi}\,(x))^2 )}.
    \end{equation}
   The working in \cite{Sm70} then implies that (\ref{1.5f}) plays the role of the scaling function $F$ in (\ref{1.5d}) for the weak
   disorder expansion of the density of states near $\omega = 2$ for Dyson's type I chain.
   Also of interest is the fact that for large $x$ the leading asymptotics of (\ref{1.5f}) is $x^{1/2} e^{- 4 x^{3/2}/3}$, where the
   particular functional form of the exponential is known in the theory of disordered systems as a Lifshitz tail; see \cite[\S 7.1]{CT16}.
   \end{remark}
   
   \section{Inhomogeneous Type I chains related to Gaussian anti-symmetric matrices}\label{S4}
   Let $X$ be a real $N \times N$ standard Gaussian matrix. The corresponding symmetric matrix ${1 \over 2} (X + X^T)$ is
   said to be a member of the Gaussian orthogonal ensemble; see e.g.~\cite[\S 1.1]{Fo10}. This ensemble relates to the
   broader theme of disordered chains and tight binding Hamiltonians  through the property
   that it permits a similarity transformation (using Householder reflection matrices) to a tridiagonal matrix with entries on
   and above the diagonal again independent \cite{Tr84}. On the diagonal they are unchanged, each being given by a standard
   Gaussian. On the sub-diagonal directly above the diagonal they are distributed by $(\tilde{\chi}_{N - 1}, \tilde{\chi}_{N - 2},\dots. \tilde{\chi}_{1})$,
   where $\tilde{\chi}_k$ denotes the square root of the gamma distribution $\Gamma[k/2,1]$. Moreover, with the latter generalised to
  $(\tilde{\chi}_{(N - 1)\beta}, \tilde{\chi}_{(N - 2)\beta},\dots. \tilde{\chi}_{\beta})$, where  $\beta > 0$ is a parameter, it was shown in 
  \cite{DE02} that the eigenvalue PDF can be explicitly computed, and is proportional to
  $$
  \prod_{l=1}^N e^{- x_l^2/2} \prod_{1 \le j < k \le N} | x_k - x_j|^\beta.
  $$
  This is the definition of the  Gaussian $\beta$-ensemble. 
  
  As discussed in \cite{BFS07}, this class of random tridiagonal  matrices is of interest from the viewpoint of
  stochastic  Schr\"odinger operators in one-dimension with a random potential decaying as $|x|^{-\alpha}$. 
  It is known that the exponent $\alpha$ equalling $1/2$ separates localised and extended states,
 and it turns out that the random tridiagonal matrices giving rise to the Gaussian
  $\beta$-ensemble corresponds to this critical case.
  
  An anti-symmetric Hermitian matrix can be formed out of a real Gaussian matrix $X$
  by forming $i$ times ${1 \over 2} (X - X^T)$. It is known \cite{DF10} that reduction of the latter
  to an anti-symmetric 
  tridiagonal form using Householder transformations gives for the entries directly above the diagonal
  the same distribution as in the symmetric case, $(\tilde{\chi}_{N - 1}, \tilde{\chi}_{N - 2},\dots. \tilde{\chi}_{1})$.
  Furthermore, as a generalisation, if an anti-symmetric tridiagonal matrix is constructed with entries directly
  above the diagonal distributed by  
   \begin{equation}\label{A1+}   
  (\tilde{\chi}_{(N - 1)\beta/2}, \tilde{\chi}_{\beta(N - 2)/2},\dots, \tilde{\chi}_{\beta}),
   \end{equation}
  it was shown in \cite{DF10} that the eigenvalue PDF can be explicitly determined. The precise functional
  form depends on the parity of $N$. Replacing $N$ by $2N+1$ so the size of the matrix is odd, there is one
  zero eigenvalue, with the remaining eigenvalues coming in pairs $\{ \pm i x_j \}_{j=1}^N$, $x_j > 0$,
  and their squares $x_j^2 =: y_j$
  distributed according to the PDF proportional to 
   \begin{equation}\label{A1}   
   \prod_{l=1}^N y_l^{3\beta/4 - 1} e^{- y_l} \prod_{1 \le j < k \le N} | y_k - y_j|^\beta.
   \end{equation} 
   We remark that up to scaling, with $\beta = \alpha$ the distribution $ \tilde{\chi}_{\beta}$ in the final entry
   of (\ref{A1+}) is precisely that which underlies the exactly solvable case of a Type I chain identified
   by Dyson --- recall (\ref{57}).
   
   The PDF (\ref{A1}) is an example of the Laguerre $\beta$-ensemble; see e.g.~\cite[\S 3.10]{Fo10}.
   After scaling the squared singular values 
  $
   y_j \mapsto \beta y_j/2 
 $
 the density of states corresponding to (\ref{A1}) has for large $N$ a Mar\u{c}enko-Pastur functional form.
 The latter    is independent of $\beta$, and is given by  (see e.g.~\cite[\S 3.4.1]{Fo10})
     \begin{equation}\label{A3} 
    D(\mu) =  {2 \over \pi \mu^{1/2}} (1 - \mu)^{1/2},
 \end{equation}      
  where $\mu = y/4N$, supported on $0 < Y <1$; cf.~(\ref{38}). 
  
  Another limiting procedure is possible. Following \cite{ABMV13} (see also \cite{TT19,Ma20,MP20})
  set $\beta = c/N$, then take $N \to \infty$. 
  Note from (\ref{A1+}) that all elements in the top $k \times k$ ($k$ fixed) sub-block have the same
  distribution $ \tilde{\chi}_{c/2}$, which with $\alpha = c/2$ (and up to a scaling) agrees with 
  the distribution specifying Dyson's exactly
  solvable Type I chain.
  In this regime  we know from \cite[Eq.~(3.49) with $\lambda /2 = \mu$]{ABMV13} that
   \begin{equation}\label{A4} 
    D(\mu) =  {1 \over   \Gamma(c) \Gamma(c+1)} {1 \over |W_{-c + 1/2,0}(-\mu) |^2},
 \end{equation}    
  where $W_{\kappa,\mu}(z)$ denotes the Whittaker function in usual notation.
  From the known $z \to 0$ asymptotics of the latter \cite[Eq.~13.14.99]{DLMF}
  it follows that for $\mu \to 0^+$
  \begin{equation}\label{A5}  
 D(\mu) \sim  {c \over \mu} {1 \over (\log \mu)^2}.
 \end{equation}
 In comparison, it follows by differentiating (\ref{72}) that for Dyson's exactly solvable
 Type I chain, and more generally to all chains in this class where the distribution has finite
 second moment (recall the discussion in \S \ref{3.3}), that for $\mu \to 0^+$
  \begin{equation}\label{A5}  
 D(\mu) \sim {\tilde{c} \over \mu}  {1 \over (\log \mu)^3},
 \end{equation} 
 for some proportionality $\tilde{c}$. Thus up to a factor of $1/ \log \mu$, for $\mu \to 0^+$ the
 density $D(\mu)$ has a functional form characteristic of a Dyson Type I chain with all random
 variables drawn from the same distribution.
 
 \begin{remark}
 With the Whittaker function in (\ref{A4}) expressed in terms of the Kummer function, and
 up to a factor of $\mu/c$, the RHS is known in the context of Dyson's exactly solvable Type I
 chain, being obtained in a calculation of a particular spectral density (in distinction to the density of
 states) \cite[second last displayed equation in \S 7.2 with $q=1$, $p=c$]{CT16}.
 \end{remark}
   
    \subsection*{Acknowledgements}
	This research is part of the program of study supported
	by the Australian Research Council Centre of Excellence ACEMS,
	and the Discovery Project grant DP210102887. I thank M.~Sodin for
	providing the references \cite{Pa73}, \cite{RB60}, K.~Borovkov for the English translation
	of the bibliographical details of the latter, and Ch.~Texier for helpful remarks on the
	first draft.

\providecommand{\bysame}{\leavevmode\hbox to3em{\hrulefill}\thinspace}
\providecommand{\MR}{\relax\ifhmode\unskip\space\fi MR }
\providecommand{\MRhref}[2]{%
  \href{http://www.ams.org/mathscinet-getitem?mr=#1}{#2}
}
\providecommand{\href}[2]{#2}

\end{document}